\documentclass[aps,prc,groupedaddress,twocolumn,showpacs,amsmath,amssymb,floatfix]{revtex4}
\usepackage{graphics}
\usepackage{dcolumn}
\usepackage{longtable}

\begin{document}
\title{Consistent account of deuteron-induced reactions on $^{nat}$Cr up to 60 MeV}

\author{E.~\v Sime\v ckov\'a$^{1}$\footnote{simeckova@ujf.cas.cz},
M.~Avrigeanu$^{2}$\footnote{marilena.avrigeanu@nipne.ro},
U.~Fischer$^3$,
J.~Mr\'azek$^1$,
J.~Novak$^1$,
M.~\v Stef\'anik$^1$,
C.~Costache$^2$,
V.~Avrigeanu$^2$,
}

\affiliation{$^1$Nuclear Physics Institute CAS, 25068 \v Re\v z, Czech Republic}

\affiliation{$^2$Horia Hulubei National Institute for Physics and Nuclear Engineering, P.O. Box MG-6, 077125 Bucharest-Magurele, Romania}

\affiliation{$^3$Euratom/FZK Fusion Association, Karlsruhe Institute of Technology (KIT), Hermann-von-Helmholtz-Platz, 1, 76344 Eggenstein-Leopoldshafen, Germany}

\date{\today}

\begin{abstract}
\noindent
{\bf Background:}
Specific noncompound processes as breakup (BU) and distinct direct reactions (DR) make the deuteron-induced reactions different from reactions with other incident particles. 
Significant discrepancies with measured cross sections were provided by taking into account only the pre-equilibrium emission (PE) and compound-nucleus (CN) mechanisms while microscopic calculation of inclusive BU and DR cross sections is still investigated numerically. Thus, reaction cross sections recommended most recently for high-priority elements are still based on data fit.

\smallskip

\noindent
{\bf Purpose:}
Accurate new measurements of low-energy deuteron-induced reaction cross sections for natural Cr target can enhance the related database and the opportunity for an unitary and consistent account of the involved reaction mechanisms. 

\smallskip

\noindent
{\bf Methods:}
The activation cross sections of $^{51,52,54}$Mn, $^{51}$Cr, and $^{48}$V nuclei for deuterons incident on natural Cr at energies up to 20 MeV, were measured by the stacked-foil technique and high resolution gamma spectrometry using U-120M cyclotron of the Center of Accelerators and Nuclear Analytical Methods (CANAM) of the Nuclear Physics Institute of the Czech Academy of Sciences (NPI CAS). They as well as formerly available data for deuteron interactions with Cr isotopes up to 60 MeV are the object of an extended analysis of all processes from elastic scattering until the evaporation from fully equilibrated compound system, but with a particular attention given to the BU and DR mechanisms. 

\smallskip

\noindent
{\bf Results:}
The new measured activation excitation functions proved essential for the enrichment of the deuteron database, while the theoretical analysis of all available data strengthens for the first time their consistent account provided that (i) a suitable BU and DR assessment is completed by (ii) the assumption of PE and CN contributions corrected for decrease of the total-reaction cross section due to the leakage of the initial deuteron flux towards BU and DR processes. 

\smallskip

\noindent
{\bf Conclusions:}
The suitable description of nuclear mechanisms involved within deuteron-induced reactions on chromium, taking into account especially the BU and DR direct processes, is validated by an overall agreement of the calculated and measured cross sections including particularly the new experimental data at low energies.
\end{abstract}


\maketitle

\section{Introduction}
\label{Sec1}

Specific noncompound processes as breakup (BU) and distinct direct reactions (DR) make the deuteron-induced reactions different from reactions with other incident particles. 
The deuteron interaction with low and medium mass target nuclei and incident energies below and around the Coulomb barrier proceeds largely through stripping and pick-up DR mechanisms, while pre-equilibrium emission (PE) and evaporation from fully equilibrated compound nucleus (CN) become important at higher energies (e.g., \cite{ma17} and Refs. therein).
Moreover, the deuteron BU is quite important along the whole energy range. 
Thus, significant discrepancies with measured cross sections are due to account of only PE and CN mechanisms while microscopic calculation of inclusive BU and DR cross sections is still investigated numerically (e.g., \cite{lei18} and Refs. therein). 
On the other hand, the sparce deuteron-breakup experimental data systematics \cite{pamp78,wu79,mats80,klein81,must87} related to the high complexity of the breakup mechanism has constrained so far a comprehensive analysis of the deuteron interactions within wide ranges of target nuclei and incident energies.
Thus, reaction cross sections recommended most recently for high-priority elements are still based on data fit, e.g. by various-order Pade approximations \cite{herm18}, with so low predictive power and apart from nuclear modeling advance.

Consequently, complementary measurements and model calculations are necessary in order to meet the demands of several on-going strategic research programmes of international large-scale facilities \cite{iter,ifmif,nfs} and databases \cite{FENDL}.
The present work aims both to enhance the database of deuteron-induced reactions on natural chromium, up to 20 MeV, and continue a series of recent analyzes \cite{BU,Ald,Cud,Nbd,Fed,Nid} looking for the consistent inclusion of the deuteron BU contribution within activation cross-section account. 
The experimental setup as well the new measured data are described in Sec.~\ref{exp}. The assessment of the models involved within present work concerns firstly an energy-dependent optical potential for deuterons on Cr isotopes in Sec.~\ref{omp}. Deuteron BU account is reviewed with reference to the corresponding activation cross sections of Cr isotopes in Sec.~\ref{BU}, while the DR analysis using the computer code FRESCO \cite{FRESCO} is described in Sec.~\ref{DR}. The main points of the PE and CN contributions obtained using the code TALYS-1.9 \cite{talys} are given in Sec.~\ref{PE+CN}. The comparison of the measured and calculated deuteron activation cross sections of Cr stable isotopes and natural Cr is discussed in Sec.~\ref{Activation}, including the TENDL-2017 evaluated data \cite{TENDL}, while conclusions of this work are given in Sec.~\ref{Sum}.

\section{Measurements} \label{exp}

The irradiation was carried out on CANAM infrastructure \cite{CANAM} of NPI CAS using an external deuteron beam of the variable-energy cyclotron U--120M operating in the negative-ion mode. The beam was extracted using a stripping-foil extractor and was delivered to the reaction chamber through a beam line consisting of one dipole and two quadrupole magnets. The mean beam energy was determined with an accuracy of 1\%, with full width at half maximum (FWHM) of 1.8\%.

\begin{figure} [!b]
\resizebox{0.55\columnwidth}{!}{\includegraphics{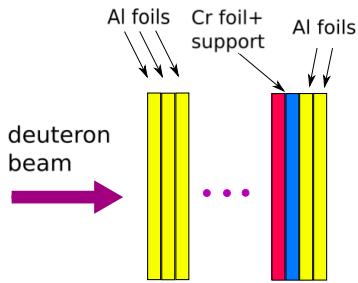}}
\caption{(Color online) The schematic picture of the foil stacks used in the experiment. The Cr foil was obtained on a plastic support.}
\label{Stacks2}
\end{figure}

The activation cross sections were measured by the stacked-foil technique, where the measured foils are interleaved by monitoring/degrading foils. 
The Cr foils (Goodfellow product - 99.99\% purity, 25 $\mu$m declared thickness) were delivered on a permanent polyester support. Since we do not have an information over the properties (thickness and composition) of the support, we cannot reliably calculate beam energy losses after the Cr. Therefore we decided to use only one Cr foil per stack (Fig.~\ref{Stacks2}) and the Al foils (Goodfellow product - 99.99\% purity, 50 $\mu$m declared thickness) were used to degrade the beam energy and for an additional monitoring as well.

The collimated deuteron beam impinged the stack of foils placed in a cooled reaction chamber that served also as a Faraday-cup. Accuracy of the current/charge measurement was 5\%. Each run was last 5 - 20 min with mean current 0.15 - 0.49 $\mu$A. The initial energies were 20 MeV and 12.5 MeV. 
The mean energy, the energy thickness and the energy spread in each foil were simulated by SRIM 2008 package \cite{SRIM}. 

Natural chrome consists of four stable isotopes – $^{50}$Cr (4.345\%), $^{52}$Cr (83.789\%), $^{53}$Cr(9.501\%) and $^{54}$Cr (2.365\%) – which leads to many open reaction channels. 

\begin{table} 
\caption{\label{tab:decay} Half-lives, main gamma lines and their intensities \cite{chu99} of the isotopes observed from irradiated Cr foils.}
\begin{tabular}{cccc}\\
\hline \hline
    Isotope   & T$_{1/2}$ & E$_{\gamma}$(keV) & I$_{\gamma}$(\%) \\
\hline
$^{54}$Mn     & 312.3 day  & 834.25            & 99.98 \\
$^{52}$Mn     & 5.591 day  & 1434.07           & 100   \\
	      &            & 935.54            & 94.5  \\
	      &            & 744.23            & 90.0  \\
$^{52}$Mn$^m$ & 21.1 min   & 1434.07           & 98.3  \\
$^{51}$Mn     & 46.2 min   & 749.07            & 0.26  \\
$^{51}$Cr     & 27.7025 day& 320.08            & 10    \\
$^{48}$V      & 15.9735 day& 983.52            & 99.98 \\
              &            & 1312.96           & 97.5  \\
\hline \hline
\end{tabular}
\end{table}
 
\begingroup
\squeezetable
\begin{table*} 
\caption{\label{tab:cs} Measured reaction cross sections (mb) for deuterons incident on natural chromium. The uncertainties are given in parentheses, in units of the last digit.}
\begin{ruledtabular}
\begin{tabular}{ccccccc}\\
 Energy  & \multicolumn{6}{c}{Reaction}\\
           \cline{2-7}
   (MeV) & \rotatebox{00}{Cr$(d,xn)^{54}$Mn}
	       & \rotatebox{00}{Cr$(d,xn)^{52}$Mn$^g$}
				 & \rotatebox{00}{Cr$(d,xn)^{52}$Mn$^m$}
	       & \rotatebox{00}{Cr$(d,xn)^{51}$Mn}
				 & \rotatebox{00}{Cr$(d,x)^{51}$Cr}
				 & \rotatebox{00}{Cr$(d,x)^{48}$V}\\
\hline
19.45 (27)&17.59 (111)&143.17 (831)& 80.8    (82)&          &41.93 (459)&0.661  (40)\\
18.46 (29)&23.22 (144)&161.03 (972)&112.45 (1301)&          &28.54 (303)&0.817  (61)\\
17.45 (29)&25.83 (149)&154.59 (889)&118.9    (74)&          &18.65 (194)&1.048  (62)\\
15.83 (31)&29.06 (240)&146.88 (859)&114.9   (105)&          &14.78 (156)&1.297  (76)\\
13.56 (35)&29.18 (193)&119.35 (691)& 92.65 (1036)&          &16.22 (172)&2.087 (124)\\
11.26 (40)&30.36 (180)& 60.86 (351)& 49.05  (341)&          &19.31 (205)&2.521 (149)\\
10.00 (43)&33.49 (196)& 45.25 (264)& 32.32  (304)&          &21.21 (221)&2.617 (153)\\
 8.54 (49)&36.61 (234)&  9.30  (54)&  4.50   (63)&5.13  (61)&25.75 (268)&2.282 (132)\\
 8.29 (50)&31.8  (190)&  2.95  (17)&  1.31   (15)&5.50 (114)&26.00 (271)&1.850 (111)\\
 5.08 (69)&25.86 (151)&            &             &6.71  (87)&31.30 (329)&1.005  (60)\\
 4.30 (77)&33.22 (193)&            &             &5.99  (83)&27.76 (289)&0.774  (45)\\
 3.91 (82)&31.09 (182)&            &             &4.33  (59)&20.92 (218)&0.445  (26)\\
\end{tabular}
\end{ruledtabular}
\end{table*}
\endgroup

The gamma-rays from the irradiated foils were measured repeatedly by two calibrated high-purity germanium (HPGe) detectors of 50\% efficiency and FWHM of 1.8 keV at 1.3 MeV. Experimental reaction rates were calculated from the specific activities at the end of the irradiation and corrected for the decay during the irradiation using the charge measurement and Cr foil characteristics as well. The measurement with different cooling times lasted up to 100 days after irradiation. The decay data of the isotopes observed from irradiated Cr foils \cite{chu99} are given in Table~\ref{tab:decay}.

The experimental cross sections of the Cr$(d,x)^{54}$Mn, Cr$(d,x)^{52}$Mn$^g$, Cr$(d,x)^{52}$Mn$^m$, Cr$(d,x)^{51}$Mn, Cr$(d,x)^{51}$Cr and Cr$(d,x)^{48}$V reactions are given in Table~\ref{tab:cs}.
The energy errors take into account the energy thickness of each foil and the initial-energy spread error. Cross-section errors are composed of statistical errors in activity determination and systematical errors of charge measurement uncertainty ($\sim$5\%), foil thickness uncertainty (2\%) and uncertainty of HPGe detector efficiency determination (2\%). 
The measured cross sections are in a good agreement with recent data and will be discussed in Sec.~\ref{Activation}.

The $^{52}$Mn (T$_{1/2}$ = 5.591 day) production cross sections are actually the $^{52}$Mn$^g$ ones, only 1.7\% (this effect is hidden within experimental uncertainties) is fed from the isometric state (T$_{1/2}$ = 21.1 min). Both ground and isomeric states decay mainly through 1434.07 keV gamma-line. Using minimization procedure \cite{min} the $^{52}$Mn$^{m}$ cross section was determined from the time dependence of the 1434.07 keV gamma-line activity.

The gamma-lines from the decay of $^{51}$Mn (T$_{1/2}$ = 46.2 min) are very weak (0.26 \% is the strongest one), so we can determine cross section only in few cases at the maximum.
The cross section for the Cr$(d,xn)^{51}$Cr is in fact a cumulative one as $^{51}$Mn decays to $^{51}$Cr.

\section{Nuclear model analysis}
\label{models}
\subsection{Optical potential assessment}
\label{omp}

The optical model potential (OMP) parameters, which are obtained by fit of the deuteron elastic-scattering angular distributions, are then used within the calculation of all deuteron reaction cross sections. Thus, the simultaneous analysis of elastic-scattering and activation data appears essential for nuclear model calculations using a consistent input parameter set \cite{Ald,Cud,Nbd,Fed,Nid}. 

\begin{figure*} [!hb]
\resizebox{2.06\columnwidth}{!}{\includegraphics{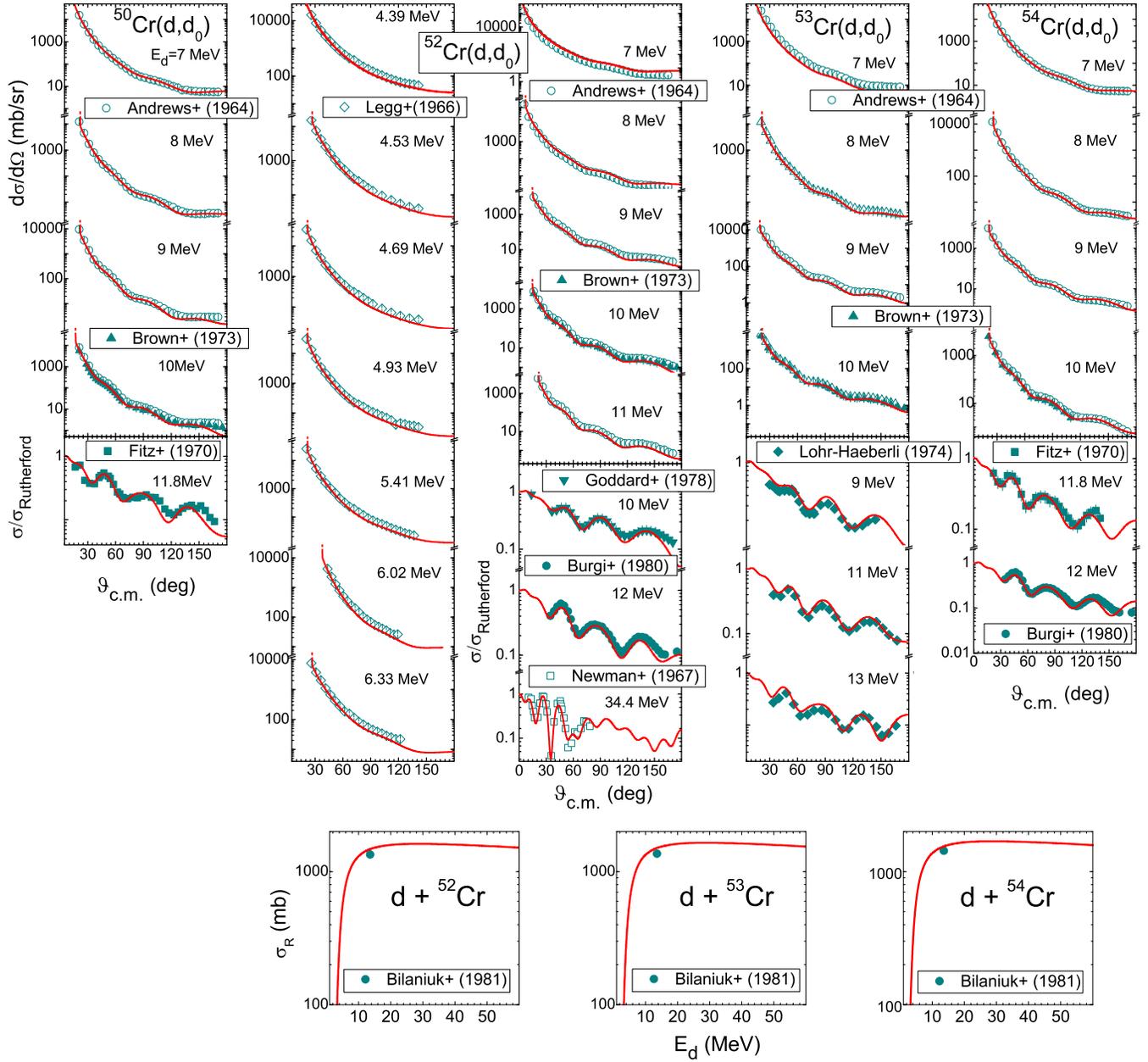}}
\caption{(Color online) Comparison of (top) measured 
\cite{andrews,legg,newman,fitz,brown,lh,goddard,burgi,exfor}
and calculated elastic-scattering angular distributions of deuterons on $^{50,52,53,54}$Cr at energies from $\sim$5 to $\sim$34 MeV, using the global OMP of Daehnick {\it et al.} \cite{dah}, and (bottom) measured \cite{exfor,bilaniuk} and similarly calculated total-reaction cross sections for deuterons on $^{52,53,54}$Cr from 3 to 60 MeV.}
\label{Cr_ad_el2}
\end{figure*}

Because the Daehnick {\it et al.} \cite{dah} OMP was established by analysis of the angular distributions of elastic-scattered deuterons on $^{52,53}$Cr isotopes too, covering almost the whole deuteron incident energies of interest for the present analysis, it was also involved within present analysis. 
The comparison of the experimental elastic-scattering angular distributions for $^{50,52-54}$Cr \cite{andrews,legg,newman,fitz,brown,lh,goddard,burgi,exfor}, at deuteron energies from $\sim$5 MeV towards 34 MeV, and the calculated values obtained by using this OMP and the computer code SCAT2 \cite{SCAT2} is shown in Fig.~\ref{Cr_ad_el2}. 
The same comparison for the sole measured total-reaction cross sections $\sigma_{R}$ of deuterons on $^{52-54}$Cr isotopes \cite{bilaniuk} is also shown.  

On the whole, the good description of the measured elastic-scattering angular distributions and the suitable account of the available $\sigma_R$ data supported well the OMP of Daehnick {\it et al.} \cite{dah} for the further use in the calculation of the activation cross sections for the deuteron interaction with Cr isotopes.

\subsection{Deuteron breakup}
\label{BU}

The physical picture of the deuteron breakup in the Coulomb and nuclear fields of the target nucleus being recently emphasized \cite{ma17,BU,Ald,Cud,Nbd,Fed,Nid,Pad,ma15}, only particular points are mentioned here. They concern the two distinct BU processes, i.e. the elastic breakup (EB) in which the target nucleus remains in its ground state and none of the deuteron constituents interacts with it, and the inelastic breakup or breakup fusion (BF), where one of these deuteron constituents interacts nonelastically with this nucleus. 
Empirical parametrizations \cite{ma17,BU} of both the total (EB+BF) $breakup$-$nucleon$ emission $f_{BU}^{n/p}$ = $\sigma^{n/p}_{BU}/\sigma_R$ and EB $f_{EB}$=$\sigma_{EB}/\sigma_R$ fractions  provide finally the BU cross sections under the assumption of equal neutron- and proton-emission breakup cross sections. The BF fraction is given by the difference $f_{BF}^{n/p}$=$f_{BU}^{n/p}$-$f_{EB}$ as well. While the experimental systematics of the total breakup $proton$-emission fraction covers a large range of target nuclei, from $^{27}$Al to $^{232}$Th and incident deuteron energies from 15 to 80 MeV, that of the elastic breakup fraction covers an energy range up to 30 MeV. Because of that, the correctness of the extrapolation of corresponding parametrization has been checked by comparison \cite{maCDCC} with results of the microscopic CDCC method \cite{CDCC1,CDCC2,CDCC3,CDCC4}. 
Thus, a normalization factor has been introduced for the extrapolation of $f_{EB}$ at energies beyond the available data, in agreement with the behavior of $f_{BU}^{p}$ and the CDCC calculation results \cite{ma17}.

\begin{figure} [!tbp]
\resizebox{1.00\columnwidth}{!}{\includegraphics{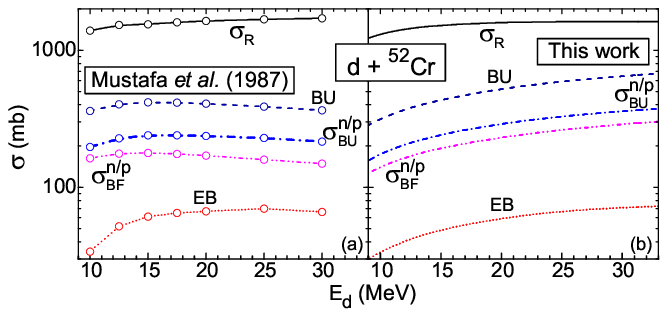}}
\caption{(Color online) Comparison of the total-reaction cross sections (solid curves) as well as cross sections of total BU (dashed curves), total BU $nucleon$-emission (dash-dotted curves), inelastic-breakup nucleon emission (dash-dot-dotted curves), and EB (dotted curves) provided by (a) microscopic predictions of Mustafa {\it et al.} \cite{must87}, and (b) presently involved parametrization \cite{ma17,BU}.}
\label{52Cr_BU_comp}
\end{figure}
\begin{figure*} 
\resizebox{1.85\columnwidth}{!}{\includegraphics{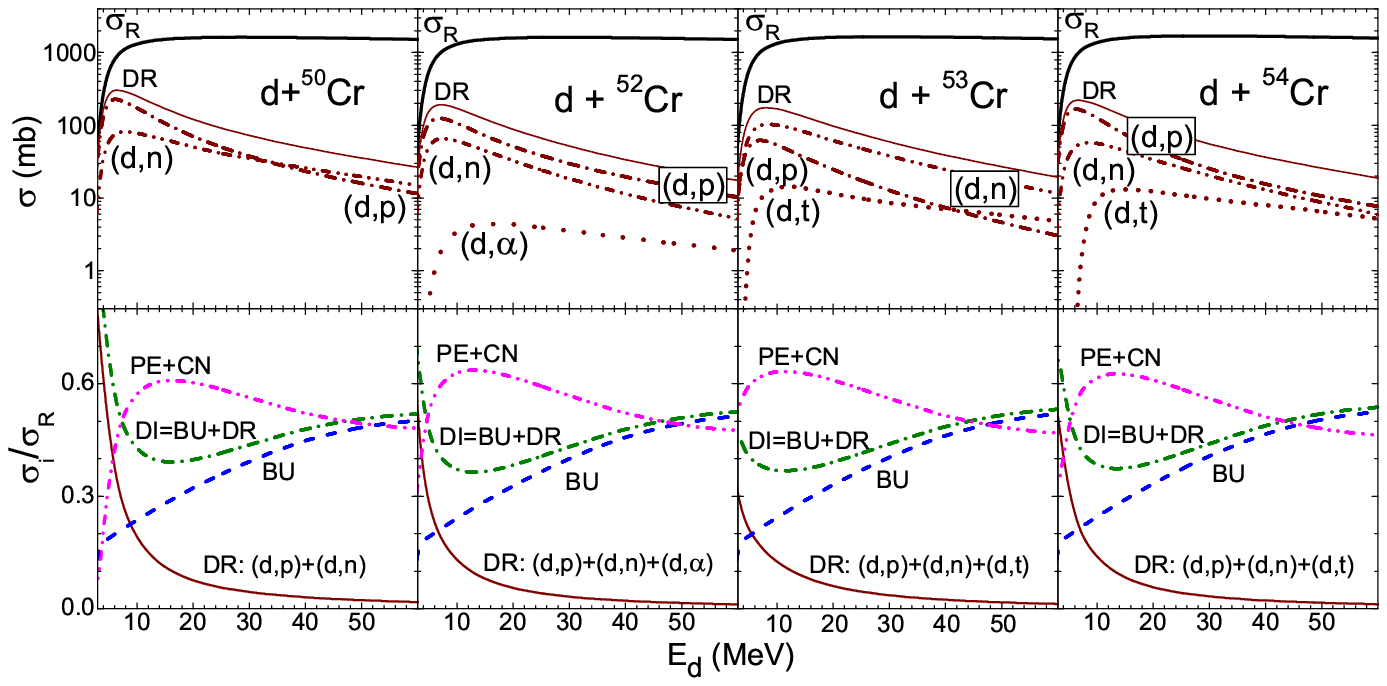}}
\caption{(Color online) (Top) Total-reaction (thick solid curves), DR (thin solid curves), stripping $(d,n)$ (dash-dot-dotted curves) and $(d,p)$ (dash-dotted curves), and pick-up $(d,t)$ (short-dotted curves) and $(d,\alpha)$ (dotted curves) reaction cross sections of deuterons on $^{50,52-54}$Cr. (Bottom) The fractions of $\sigma_R$ corresponding to the contributions of BU (dashed curves), DR (solid curves), DI (dash-dotted curves), and PE+CN (dash-dot-dotted curves) mechanisms involved in the d+$^{54}$Cr interactions (see text).}
\label{FR_Crd}
\end{figure*}

Microscopic predictions of BU, BF, and EB cross sections for deuteron interaction with $^{52}$Cr target nucleus were provided only by Mustafa {\it et al.} \cite{must87} in the frame of the distorted-wave Born approximation (DWBA) formalism with prior form interaction \cite{TU}. Equal neutron- and proton-emission BU cross sections have been assumed too. The corresponding total BU, total BU $nucleon$-emission, inelastic-breakup nucleon emission, and EB cross sections (Table I of Ref.~\cite{must87}) are compared in Fig.~\ref{52Cr_BU_comp} with the predictions of empirical parametrization \cite{ma17,BU}. The deuteron OMPs used by Mustafa {\it et al.}, i.e. that of Lohr and Haeberli \cite{lh}, for $E_d$$\leq$13 MeV, and Perey and Perey \cite{P-P} for $E_d$$>$13 MeV, provide total-reaction cross sections rather similar to the present work while the BU cross-section differences are obvious and demand additional comments. 

The $\sigma^{n/p}_{BU}$ values obtained by Mustafa {\it et al.} decrease with the incident-energy increase above $E_{d}\sim$15 MeV. This trend is determined by $\sigma^{n/p}_{BF}$ which is not compensated by the weak $\sigma_{EB}$. The latter is yet increasing slowly with energy up to $\sim$25 MeV, but then decreases too. However, the measured BU proton-emission cross sections $\sigma^p_{BU}$ at 15 \cite{klein81}, 25.5 \cite{pamp78,klein81}, and 56 MeV \cite{mats80}, and the parametrization predictions pointed out the increase of $\sigma^p_{BU}$ with the target-nucleus mass, from $^{27}$Al up to $^{232}$Th, as well as with increasing incident energy (e.g., Fig. 2 of Ref. \cite{ma17}). These features have been evidenced also by recent microscopic breakup estimations within a  CDCC extension of the eikonal reaction theory \cite{neoh} for 56 MeV deuteron interaction with target nuclei from $^{12}$C to $^{209}$Bi. 

On the other hand, recent comparison of BU microscopic calculations \cite{lei18,lei15} on the basis of the DWBA method with prior \cite{TU} and post form interaction \cite{IAV} have pointed out missing terms of the BU differential cross sections within Tamura-Udagawa \cite{TU} theoretical frame, which lead to an underestimation of the BU cross sections. 

\begin{figure} [!hb]
\resizebox{1.0\columnwidth}{!}{\includegraphics{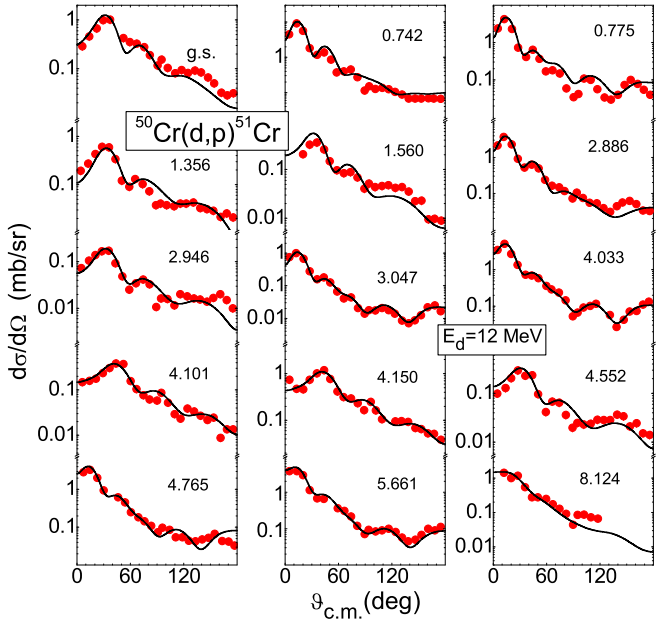}}
\caption{(Color online) Comparison of measured (solid circles) \cite{50Cr_dp} and calculated (solid curves) proton angular distributions of $^{50}$Cr$(d,p)^{51}$Cr stripping transitions to states with excitation energies in MeV, at the incident energy of 12 MeV.}
\label{50Cr_dp_fin}
\end{figure}

\begin{figure} 
\resizebox{1.0\columnwidth}{!}{\includegraphics{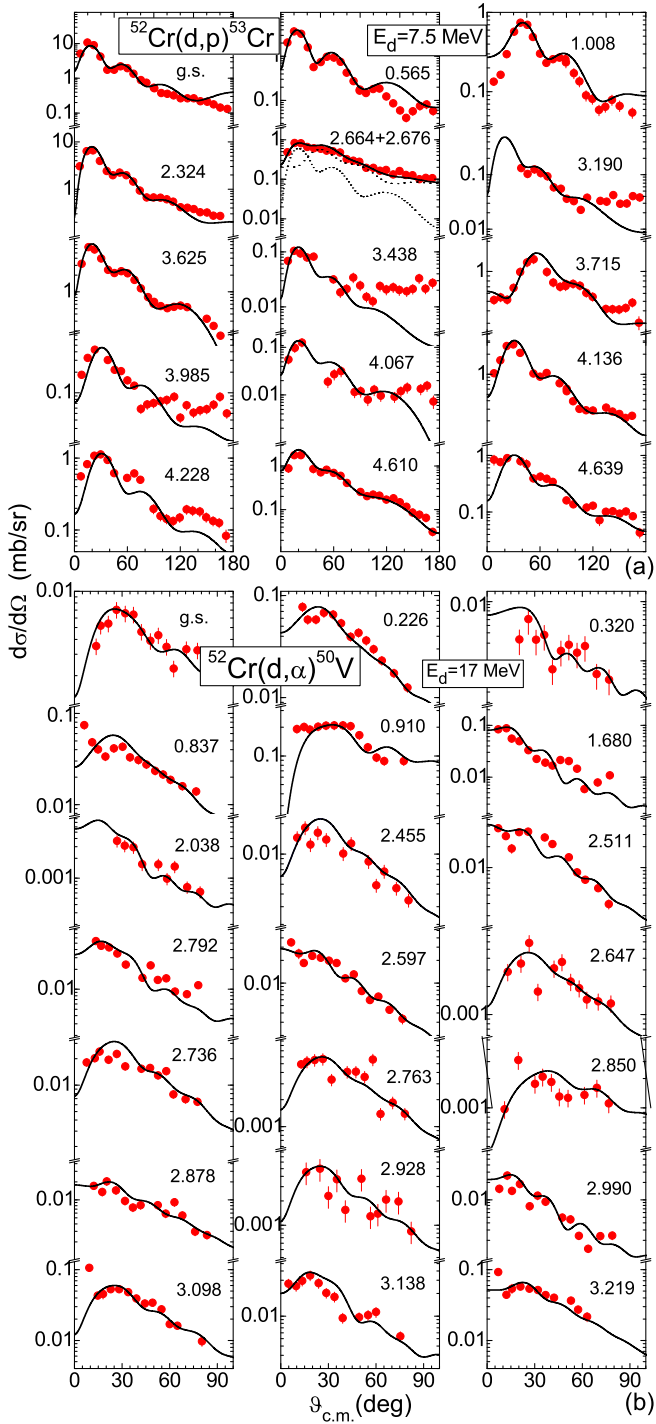}}
\caption{(Color online) As Fig.~\ref{50Cr_dp_fin} but for (a) $^{52}$Cr$(d,p)^{53}$Cr stripping \cite{52Cr_dp} and (b) $^{52}$Cr$(d,\alpha)^{50}$V pick-up \cite{52Cr_da} transitions, at the incident energies of 7.5 and 17 MeV, respectively. }
\label{52Cr_dpda_fin}
\end{figure}

\begin{figure} 
\resizebox{1.0\columnwidth}{!}{\includegraphics{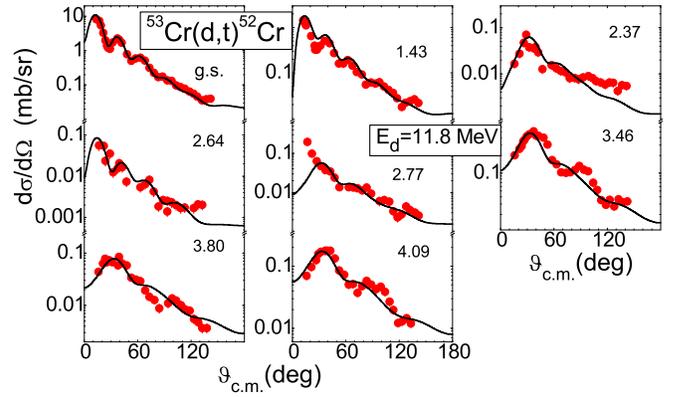}}
\caption{(Color online) As Fig.~\ref{50Cr_dp_fin} but for $^{53}$Cr$(d,t)^{52}$Cr pick-up \cite{fitz} transitions at the incident energy of 11.8 MeV.}
\label{53Cr_dt_fin}
\end{figure}

\begin{figure} 
\resizebox{1.0\columnwidth}{!}{\includegraphics{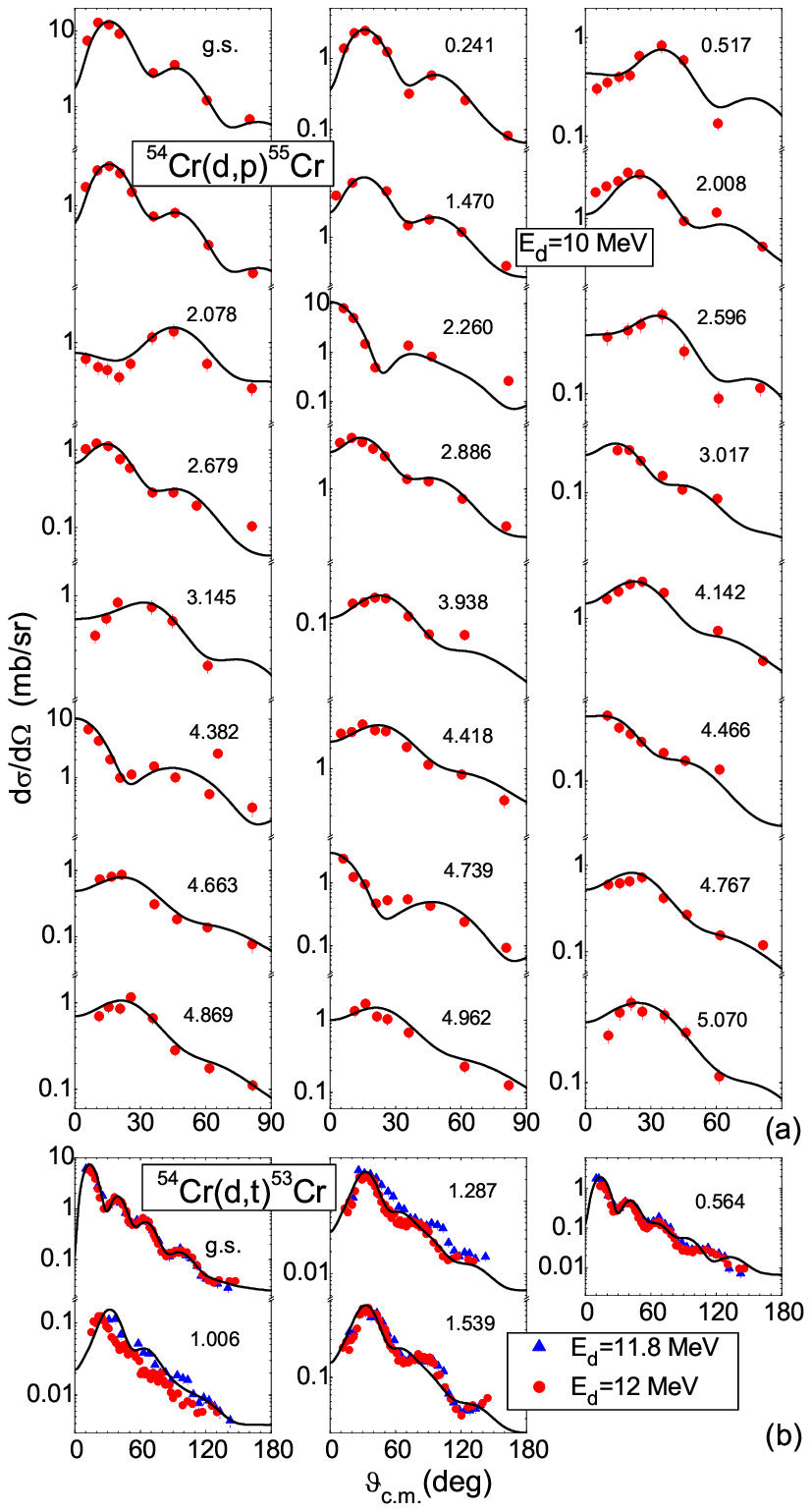}}
\caption{(Color online) As Fig.~\ref{50Cr_dp_fin} but for (a)  $^{54}$Cr$(d,p)^{55}$Cr stripping \cite{54Cr_dp}, and (b) $^{54}$Cr$(d,t)^{53}$Cr pick-up \cite{53Cr_dt,54Cr_dt} transitions, at the incident energies of 10 MeV and, respectively, 11.8 and 12 MeV.}
\label{54Cr_dpdt_fin}
\end{figure}

Nevertheless, the earlier \cite{must87} as well as actual results shown in Fig.~\ref{52Cr_BU_comp} point out the dominance of the BF component. This fact is quite important for the latter of the two BU opposite effects on the deuteron activation of the Cr isotopes too. Thus, while $\sigma_R$ that is shared among different outgoing channels, is reduced with the BU fraction $\sigma_{BU}$/$\sigma_{R}$ (bottom of Fig.~\ref{FR_Crd}), the BF component brings significant contributions to various $(d,x)$ reaction channels \cite{Ald,Cud,Nbd,Fed,Nid,Pad,ma15} through BU-nucleon interactions with the same target nucleus. The related formalism involved in the present work is described in Sec. III.B.2 of Ref. \cite{Nid}, while the BF enhancements for deuteron interaction with $^{50,52,53,54,nat}$Cr, through the $(p,x)$ and $(n,x)$ reactions, are discussed in Sec.~\ref{Activation} (Figs.~\ref{54Mn_natCr}-\ref{4746Sc_natCr}).

\subsection{Direct reactions}
\label{DR}

The assessment of transfer reaction cross sections in addition to that of BU mechanism is mandatory for the correct estimation of the final PE+CN contribution to population of various residual nuclei, in spite of poor attention given so far in deuteron activation analysis. 
The direct reactions of stripping, $(d,p)$ and $(d,n)$, and pick-up, $(d,t)$ and $(d,\alpha)$ \cite{ma15}, are quite important for the first-chance particle emission. 
However, the estimation of DR cross sections is conditioned by the available experimental spectroscopic factors, outgoing particle angular distributions, or at least the differential cross-section maximum values. 

The calculation of DR cross sections has been performed using the DWBA formalism within the FRESCO code \cite{FRESCO}. The post/prior form distorted-wave transition amplitudes for the stripping and pick-up reactions, respectively, and the finite-range interaction have been considered. The $n$-$p$ effective interaction in deuteron \cite{CDCC1} as well as $d$-$n$ effective interaction in triton \cite{triton} were assumed to have a Gaussian shape, at the same time with a Woods-Saxon shape \cite{alpha-d} of the $d$-$d$ effective interaction within the $\alpha$ particle. The transferred nucleon and deuteron bound states were generated in a Woods-Saxon real potential \cite{Ald,Cud,Nbd,Fed,Nid,ma15} while the transfer of the deuteron cluster has been taken into account for the  $(d,\alpha)$ pick-up cross section calculation. The populated discrete levels and the corresponding spectroscopic factors which have been available within the ENSDF library \cite{BNL} were used as starting input for the DWBA calculations. 

Briefly stated, the suitable description of the measured angular distributions of proton, triton, and $\alpha$-particle emission through the $(d,p)$, $(d,t)$, and $(d,\alpha)$ reactions on $^{50}$Cr (Fig.~\ref{50Cr_dp_fin}), $^{52}$Cr (Fig.~\ref{52Cr_dpda_fin}), $^{53}$Cr (Fig.~\ref{53Cr_dt_fin}), and $^{54}$Cr (Fig.~\ref{54Cr_dpdt_fin}) has supported the calculated DR excitation functions shown in the upper part of Fig.~\ref{FR_Crd}. 

A further comment should concern however the so scarce data of $(d,n)$ stripping reactions on $^{50,52,53,54}$Cr, leading to underestimated $(d,n)$ excitation functions, due to the spectroscopic factors for transitions to only few low-lying levels as well as the missing experimental angular distributions. Actually, the analysis of $^{50}$Cr$(d,n)^{51}$Mn reaction excitation function used the spectroscopic factors reported by Nilsson {\it et al.} \cite{50Cr_dn}, while the spectroscopic factors of Okorokov {\it et al.} \cite{52Cr_dn} were used to obtain the $(d,n)$ stripping excitation functions for $^{52,53,54}$Cr target nuclei shown in the upper part of Fig.~\ref{FR_Crd}. Similarly, the spectroscopic factors and the values of the angular-distribution maximum of $(d,p)$ stripping transitions to excited states of $^{54}$Cr residual nucleus reported by Bochin {\it et al.} \cite{53Cr_dp} have been involved in the calculation of the corresponding excitation function.
Therefore, only the estimation of a lower limit of the DR contribution given by the sum $\sigma_{(d,n)}$ + $\sigma_{(d,p)}$ + $\sigma_{(d,t)}$ + $\sigma_{(d,\alpha)}$, also shown in the upper part of Fig.~\ref{FR_Crd}, becomes possible.  

Nevertheless, it results that the DR contribution has a significant maximum around $E_d\sim$7 MeV mainly due to the $(d,p)$ and $(d,n)$ stripping processes. 
A slow decrease with deuteron energy follows while $\sigma_R$ reaches its maximum value and remains almost constant for $E_d$$>$ 20 MeV. This trend explains the decreasing importance of direct reactions with the deuteron energy increase while that of the  breakup mechanism becomes larger for all stable chromium isotopes, as shown in Fig.~\ref{FR_Crd}. The major role of the stripping $(d,n)$ and $(d,p)$ reactions to the activation cross sections of $^{54}$Mn, $^{51}$Mn, and $^{51}$Cr residual nuclei, for deuteron interaction with the natural Cr target, is particularly apparent in Figs.~\ref{54Mn_natCr} and \ref{51Cr_fin1}, respectively (Sec.~\ref{Activation}).

Finally, we have taken into account the deuteron total-reaction cross section that remains available for the PE+CN mechanisms, following the correction for the incident flux leakage through direct interactions (DI) of the breakup, stripping and pick-up processes, given by a reduction factor:

\begin{eqnarray}\label{eq:2}
1 - \frac{\sigma_{BU} + \sigma_{(d,n)} + \sigma_{(d,p)} + \sigma_{(d,t)} + \sigma_{(d,\alpha)}}{\sigma_R} 
     = 1 - \frac{\sigma_{DI}}{\sigma_R}.
\end{eqnarray}
The energy dependence of weighted reaction mechanism components, $\:$ $\sigma_{i}$/$\sigma_R$, $\:$ for $^{50,52,53,54}$Cr target nuclei, is also shown in Fig.~\ref{FR_Crd}. One may note firstly the high values $\sigma_{DI}$/$\sigma_R$ at lowest incident energies due to the above-mentioned maximum of the stripping excitation functions around $E_d$$\sim$7 MeV. 
The decrease of the DR component and thus of its ratio leads to a steep increase with the deuteron energy of the PE+CN weight, in spite of the BU increase. The corresponding maximum, at energies of 15-20 MeV, is followed by a decrease due to the continuous increase of BU with the incident energy. Thus, both DI and PE+CN cross sections shown in Fig.~\ref{FR_Crd} have values close to around half of $\sigma_{R}$ at energies around 60 MeV.

\subsection{Statistical PE+CN processes in deuteron-induced reactions}
\label{PE+CN}

The PE and CN statistical processes become important at the incident energies above the Coulomb barrier (Fig.~\ref{FR_Crd}). The corresponding reaction cross sections have been calculated in the frame of the PE exciton and Hauser-Feshbach formalisms using the latest TALYS-1.9 code version \cite{talys}, corrected in order to take into account the above-mentioned breakup, stripping, and pick-up components. 

The following input options of the TALYS-1.9 have been used: (a) the OMPs of Koning-Delaroche \cite{KD}, Becchetti-Greenlees \cite{BG}, and Avrigeanu {\it et al.} \cite{AHA}, for nucleons, tritons, and $\alpha$-particles, respectively, (b) the back-shifted Fermi gas (BSFG) formula for the nuclear level density, (c) no TALYS breakup contribution, since the above-mentioned BF enhancements is still under implementation in TALYS, and (d) the PE transition rates calculated by means of the corresponding OMP parameters, using the value 3 for the 'preeqmode' keyword of TALYS.

\section{RESULTS AND DISCUSSION}
\label{Activation}

\begin{figure} [t]
\resizebox{1.00\columnwidth}{!}{\includegraphics{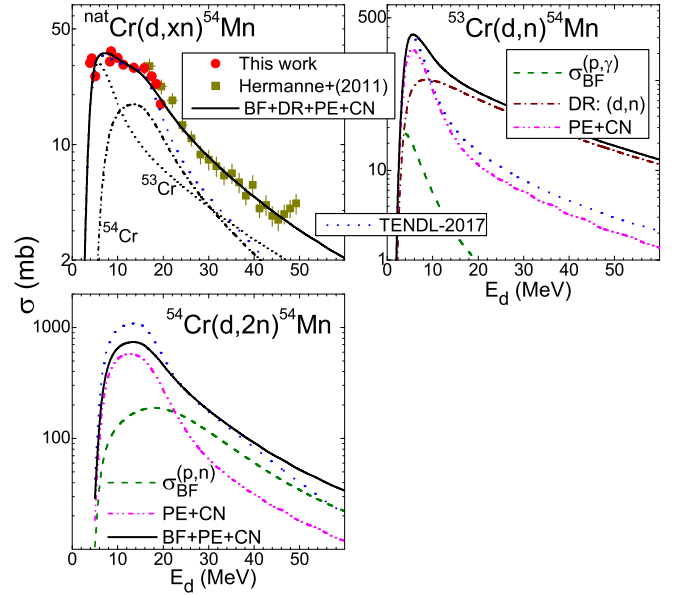}}
\caption{(Color online) Comparison of previous \cite{herm} and present (solid circles) measurements, TENDL-2017 \cite{TENDL} evaluation (dotted curves), and present calculation (solid curves) of $^{nat}$Cr$(d,xn)^{54}$Mn, $^{53}$Cr$(d,n)^{54}$Mn, and $^{54}$Cr$(d,2n)^{54}$Mn reaction cross sections, along with BF enhancement (dashed curves), stripping $(d,n)$ reaction (dash-dotted curve), and PE+CN components (dash-dot-dotted curves) corrected for DI deuteron flux leakage. 
Contributions of $^{53}$Cr (short-dashed curve) and $^{54}$Cr (short-dotted curve) isotopes to $^{nat}$Cr activation are also shown.}
\label{54Mn_natCr}
\end{figure}

\begin{figure*} 
\resizebox{2.06\columnwidth}{!}{\includegraphics{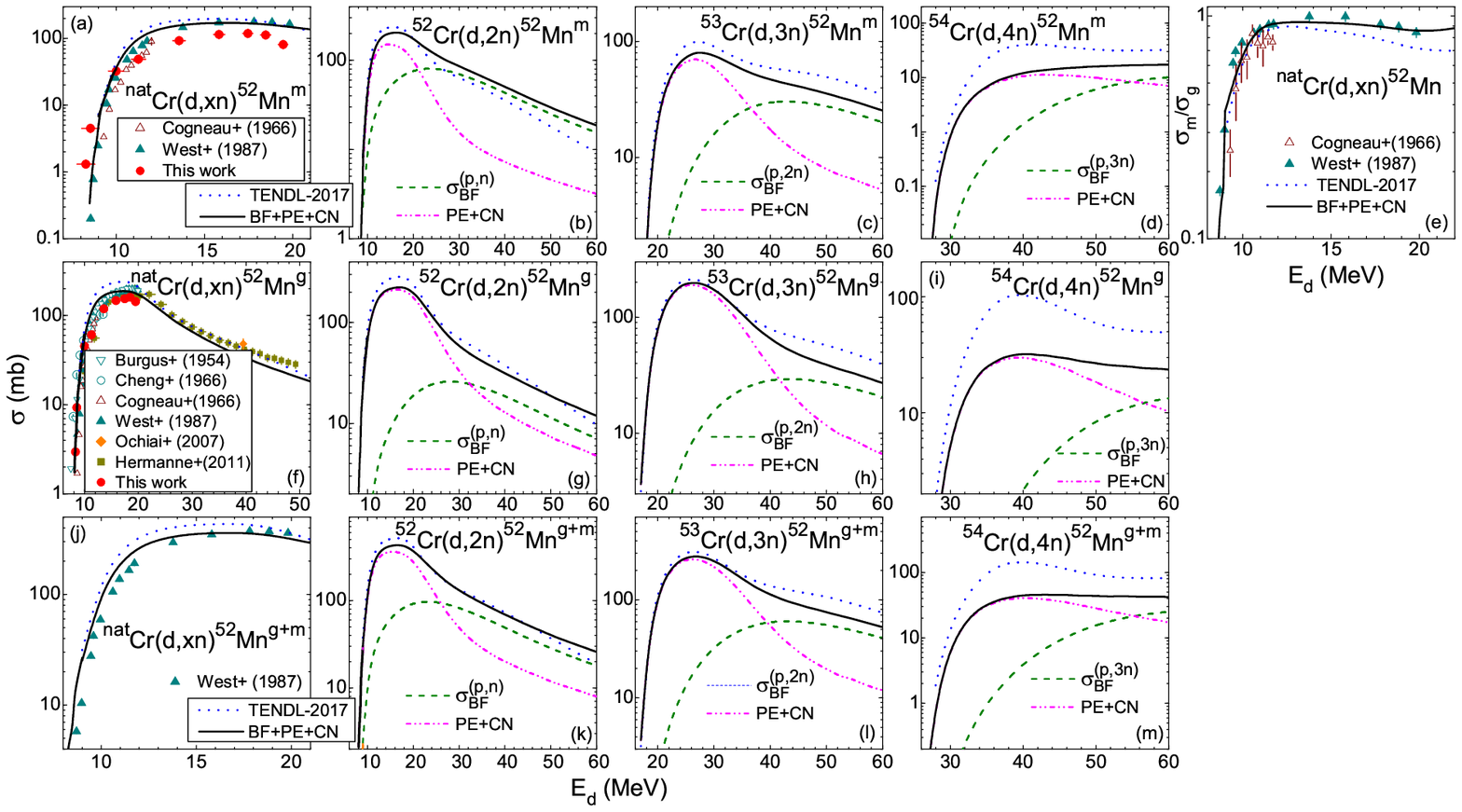}}
\caption{(Color online) As Fig.~\ref{54Mn_natCr} but for the population of (a-d) $^{52}$Mn$^m$, (f-i) $^{52}$Mn$^g$, and (j-m) $^{52}$Mn$^{g+m}$ by the corresponding $(d,xn)$ reactions on $^{nat,52,53,54}$Cr \cite{herm,ochiai,west,cheng,cogn,burg}, as well as (e) the isomeric cross section ratio $\sigma^m$/$\sigma^g$ of these reactions on $^{nat}$Cr.}
\label{52Mn_natCr} 
\end{figure*}

The excitation functions of particular residual nuclei of deuteron interaction with $^{nat}$Cr, measured in the present work (Sec.~\ref{exp}), are compared in Figs.~\ref{54Mn_natCr}--\ref{4746Sc_natCr} with the data formerly available \cite{exfor,herm,ochiai,west,cheng,burg,cogn,klein,coetz,bisconti,baron}. The isotope and mechanism detailed contributions are particularly illustrated too. 
The corresponding TENDL--2017 predictions \cite{TENDL} are also shown while a global comparison of the measured data and calculated results for $^{nat}$Cr is shown in Fig.~\ref{natCr_fin}. 
Additional comments concern several reaction types and residual nuclei as follows. 

\subsection{$(d,xn)$ reactions and $^{51,52,54}$Mn residual nuclei population}
\subsubsection {The $^{nat}$Cr$(d,xn)^{54}$Mn reaction} 

The present discussion concerns firstly the heavier residual nuclei which are populated by deuteron interactions with fewer Cr isotopes and emission of fewer nucleons. 
Thus, a more focused analysis of distinct reaction mechanisms becomes possible in the case of $^{54}$Mn nucleus. 
Unfortunately there are no data corresponding to the residual nucleus $^{54}$Mn by activation of $^{53,54}$Cr isotopes, which would represent a stronger check of the nuclear model approach. 
Therefore the ultimate model validation is provided by the comparison of the data and calculated results for the natural target, which demands for well increased accuracy of both the measurements and model analysis. 

First, one should note the major role of the stripping contribution to the $^{53}$Cr$(d,n)^{54}$Mn excitation function shown in Fig.~\ref{54Mn_natCr}. It exceeds by far the PE+CN components, increasing also the $^{54}$Mn residual-nucleus population for $^{nat}$Cr activation shown in the same figure. Moreover, there are thus provided grounds for the significant underestimation above $\sim$20 MeV of the measured data by the TENDL-2017 evaluation.

Second, the population of $^{54}$Mn residual nucleus through the $(d,2n)$ reaction is increased by the BF contribution brought by breakup protons through $(p,n)$ reaction. 
It exceeds the PE+CN components at deuteron energies above 20 MeV, leading within this energy range to a much slower decrease of this excitation function which has a steep increase above the threshold. 

On the whole, the comparison of the present results with TENDL evaluation points out the importance of the new measured cross sections around the maximum of the $^{nat}$Cr$(d,xn)^{54}$Mn activation as well as the role of breakup and stripping mechanisms to obtain the suitable description of these data. The good agreement of experimental and calculated excitation function shown in Fig.~\ref{54Mn_natCr} supports the correctness of the corresponding reaction models. 

\subsubsection{The $^{nat}$Cr$(d,xn)^{52}$Mn reaction}

A complex analysis concerns the  ground state (g.s.), isomeric state, and total populations of $^{52}$Mn residual nucleus, as well as of the isomeric cross section ratio $\sigma^m$/$\sigma^g$, as a result of deuteron interaction with $^{nat}$Cr (Fig.~\ref{52Mn_natCr}). Actually, the $2^+$ ($T_{1/2}$=21 min) isomeric state of $^{52}$Mn has only a weak 1.75\% decay branch to the $6^+$ ($T_{1/2}$=5.6 d) g.s., the rest of its decay going towards $^{52}$Cr residual nucleus. Thus it is obvious the usefulness of the new measured data for the population of $^{52}$Mn$^g$ and $^{52}$Mn$^m$ states.

The BU, PE and CN reaction mechanisms are involved in the population of these states by deuterons incident on $^{52,53,54}$Cr isotopes. 
The statistical PE+CN emission gives the largest contribution to $(d,2n)$, $(d,3n)$, and $(d,4n)$ yields only between the reaction thresholds and the excitation function maximum. 
Next, the cross sections decrease slower at higher incident energies due to the larger contributions brought by the breakup protons through $(p,n)$, $(p,2n)$, and $(p,3n)$ reactions (Fig.~\ref{52Mn_natCr}). 
The present and previous \cite{herm,ochiai,west,cheng,cogn,burg} measurements of the isomeric, g.s., and total cross sections for the population of $^{52}$Mn by deuteron interaction with $^{nat}$Cr are compared with the sum of calculated results for $^{52,53,54}$Cr isotopes in Figs.~\ref{52Mn_natCr}(a,f,j), and finally their agreement validates the account of the BF+PE+CN contributing mechanisms.

A particular analysis of the measured isomeric cross section ratio by West {\it et al.} \cite{west} and Cogneau {\it et al.} \cite{cogn} is shown in Fig.~\ref{52Mn_natCr}(e). In order to describe the experimental isomeric-ratio excitation function we found necessary to use a normalization factor of 0.75 for the spin cut-off parameter of the residual nucleus $^{52}$Mn. 
This adjustment of the width of the angular momentum distribution of the level density within the code TALYS is actually close to the ratio of the effective moment of inertia of the nucleus to the rigid-body value for $A$$\sim$50 \cite{va2001}. Thus, it seems to be an alternate choice to the option of an energy-dependent moment of inertia \cite{inertia1,inertia2,inertia3,inertia4}.

\begin{figure*} 
\resizebox{2.06\columnwidth}{!}{\includegraphics{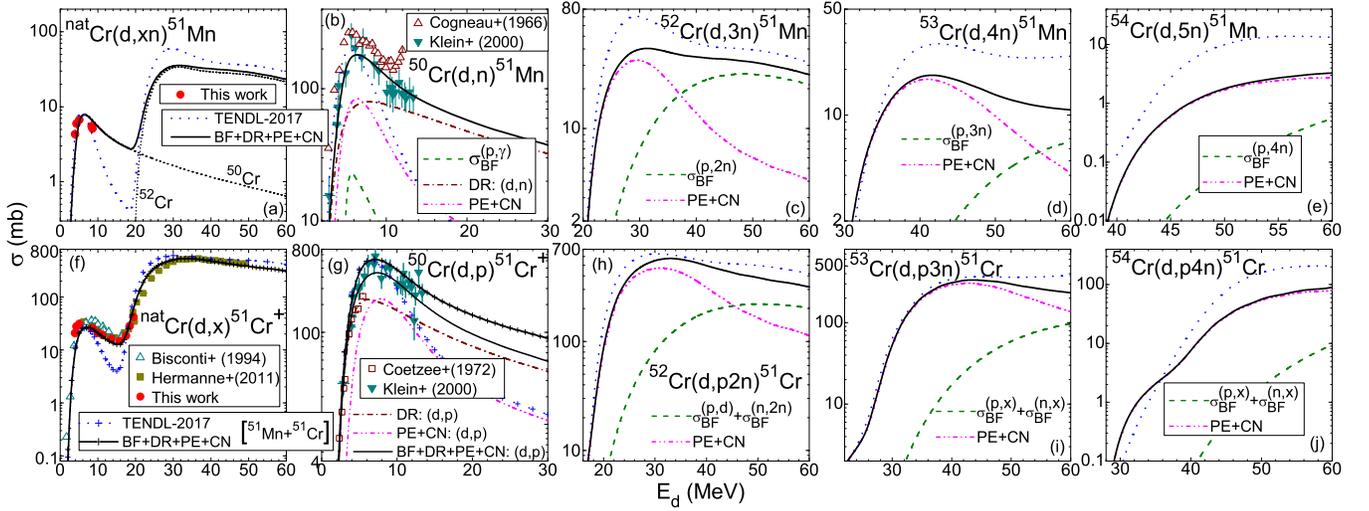}}
\caption{(Color online) As Fig.~\ref{54Mn_natCr} but for the population of (a-e) $^{51}$Mn, and (f-j) $^{51}$Cr by the corresponding $(d,xn)$ and $(d,pxn)$ reactions, respectively, on $^{nat,52,53,54}$Cr \cite{herm,cogn,klein,coetz,bisconti}, including 
(a) the contributions of $^{50}$Cr (short-dashed curve) and $^{52}$Cr (short-dotted curve) isotopes to $^{nat}$Cr activation for the former residual nucleus $^{51}$Mn, and
(f,g) $^{51}$Cr cumulative population (additional crosses) due to the decay of $^{51}$Mn residual nucleus, in comparison with the measured data for deuterons on $^{nat,50}$Cr \cite{herm,klein,coetz,bisconti} (see text).}
\label{51Cr_fin1} 
\end{figure*}

\subsubsection{The $^{nat}$Cr$(d,xn)^{51}$Mn reaction}

A very interesting case is that of the activation cross sections of $^{51}$Mn residual nucleus following the deuteron interactions with $^{nat}$Cr. Because of the energy thresholds of the corresponding $(d,3n)$, $(d,4n)$ and $(d,5n)$ reactions on $^{52,53,54}$Cr isotopes,  visible also in Figs.~\ref{51Cr_fin1}(c-e), the $^{50}$Cr$(d,n)^{51}$Mn reaction is the only one contributing to the $^{nat}$Cr$(d,xn)^{51}$Mn excitation function for incident energies $\le$19 MeV. This fact is most important particularly due to the absence of any strong contributions coming from the $\sim$84\% most abundant $^{52}$Cr isotope.

Thus it results that the stripping $(d,n)$ reaction is critical for the suitable account of both $^{50}$Cr$(d,n)^{51}$Mn and $^{nat}$Cr$(d,xn)^{51}$Mn excitation functions shown in Figs.~\ref{51Cr_fin1}(a,b), providing a distinct test of the nuclear model approach. The firstly measured cross sections for the latter reaction in the present work, in addition to only a couple of data sets already available for the former one \cite{cogn,klein}, becomes so of large interest in this respect.

Moreover, it is obvious that the suitable description of the measured excitation functions in Figs.~\ref{51Cr_fin1}(a,b) corresponds to the DR stripping account. The major importance of this reaction mechanism is proved also by the substantial underestimation of the same cross sections by the corresponding TENDL evaluation which is known to be less accurate with respect to the deuteron DR account. 

A further distinct point of $^{51}$Mn residual-nucleus population concerns the important BF enhancement following the BU-proton interactions with $^{50,52,53,54}$Cr target nuclei through $(p,xn)$ reactions as shown in Figs.~\ref{51Cr_fin1}(b-e). One may note its larger contribution versus PE+CN processes at energies above 40--50 MeV, for the target nuclei $^{52,53}$Cr.

\subsection{The $^{nat}$Cr$(d,pxn)^{51}$Cr reaction}

In this case a cumulative process should be considered in a similar way to the measured activation cross sections for the residual nucleus $^{51}$Cr in deuteron-induced reaction on $^{nat}$Fe \cite{Fed} and $^{nat}$Ni \cite{Nid}. 
The EC decay of long-lived $^{51}$Cr radionuclide ($T_{1/2}$=27.7 d) \cite{51V}, populated through the reactions $^{50,52,53,54}$Cr$(d,pxn)^{51}$Cr, is added to populations by EC decay of relatively short-lived $^{51}$Mn ($T_{1/2}$=46.2 min) \cite{51V} (see Fig. 18 of Ref.~\cite{Fed}) activated through the above-discussed reactions $^{50,52,53,54}$Cr$(d,xn)^{51}$Mn. 
However, an essential difference concerns now the simpler reaction channels involved in population of the residual nuclei $^{51}$Cr and $^{51}$Mn, which makes the related model analysis more interesting as well as more useful the corresponding new measured data in the present work.

Thus, the contributions of the $(d,pxn)$ as well as $(d,xn)$ reactions have been taken into account within the analysis of $^{nat}$Cr$(d,x)^{51}$Cr reaction data shown in Fig.~\ref{51Cr_fin1}(f). Moreover, likewise the residual-nucleus $^{51}$Mn activation at incident energies below $\sim$18 MeV, $^{51}$Cr activation at the same energies follows only the deuteron interactions with the $^{50}$Cr stable isotope, e.g. Figs.~\ref{51Cr_fin1}(h-j). 
A stripping reaction, but the $(d,p)$ one, provides also the most important contribution to this activation, with the additional feature of the cumulative population of $^{51}$Cr residual nucleus in Fig.~\ref{51Cr_fin1}(g) due to the summed $^{50}$Cr$(d,p)^{51}$Cr and $^{50}$Cr$(d,n)^{51}$Mn excitation functions. 

Moreover, the agreement of the present calculations and measured data further supports the present approach of BU+DR+PE+CN mechanisms, pointing out the important role of the stripping DR. It is the gradual decreasing slope of the cumulative $^{50}$Cr$(d,x)^{51}$Cr excitation function up to $\sim$20 MeV which then is responsible for the first maximum of the same excitation function for $^{nat}$Cr target. The new measured cross sections just within this critical energy range play thus a key role for the model analysis validation. On the other hand, the recurring TENDL-2017 underestimation proves the above-mentioned drawback of the deuteron DR account .

One may also note that significant BF enhancements are brought by BU nucleons to $^{52,53,54}$Cr$(d,x)^{51}$Cr reactions at higher energies as shown in Figs.~\ref{51Cr_fin1}(h-i)), while the contribution of BU neutrons through the $(n,\gamma)$ on the target nucleus is too weak to occur in Fig.~\ref{51Cr_fin1}(g). However, further measurements at higher energies \cite{nfs} would be most useful in this respect.

\begin{figure*} 
\resizebox{2.06\columnwidth}{!}{\includegraphics{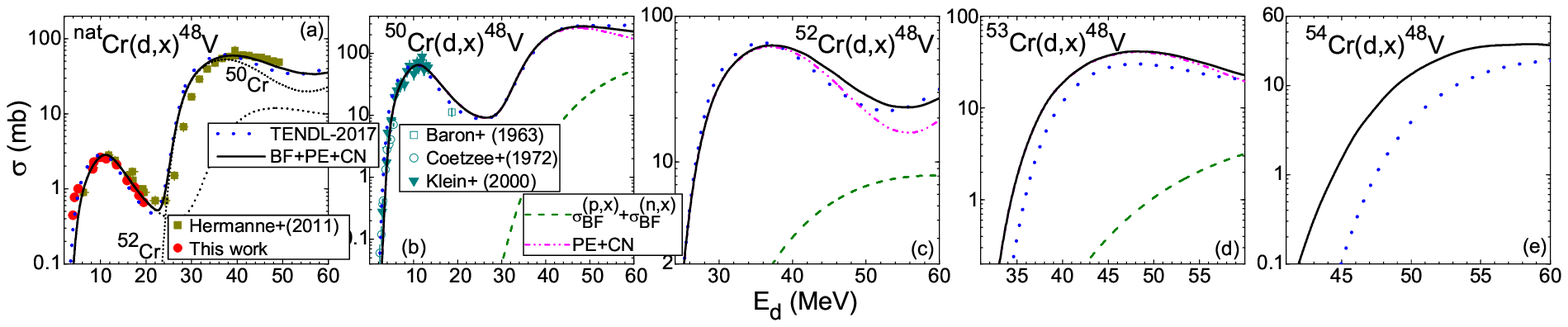}}
\caption{(Color online) As Fig.~\ref{51Cr_fin1} but for the population of (a-e) $^{48}$V by the corresponding $(d,x)$ reactions on $^{nat,52,53,54}$Cr \cite{herm,klein,coetz,baron}, including (a) the contributions of $^{50}$Cr (short-dashed curve) and $^{52}$Cr (short-dotted curve) isotopes to $^{nat}$Cr activation.}
\label{48V_natCr}
\end{figure*}

\begin{figure*} 
\resizebox{2.06\columnwidth}{!}{\includegraphics{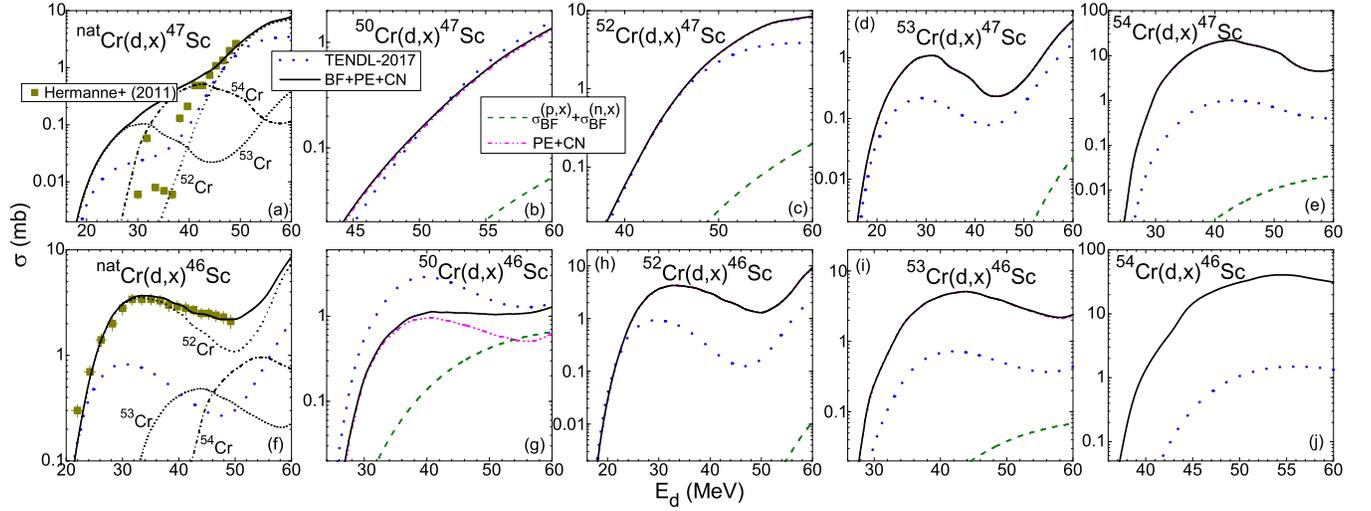}}
\caption{(Color online) As Fig.~\ref{51Cr_fin1} but for the population of (a-e) $^{47}$Sc and (f-j) $^{46}$Sc by the corresponding $(d,x)$ reactions on $^{nat,52,53,54}$Cr \cite{herm}, including (a,f) the contributions of $^{52}$Cr (short-dashed curves), $^{53}$Cr (short-dotted curves), and $^{54}$Cr (short dash-dotted curves) isotopes to $^{nat}$Cr activation.}
\label{4746Sc_natCr}
\end{figure*}

\subsection{The $(d,\alpha x)$  reactions}

\subsubsection{The $(d,\alpha xn)$  reactions}

The $(d,\alpha xn)$ reaction thresholds for Cr stable isotopes (Fig.~\ref{48V_natCr}) result in the population of the $^{48}$V nucleus by deuterons on $^{nat}$Cr up to $\sim$20 MeV only through the $(d,\alpha)$ reaction on $^{50}$Cr. The new data measured in this work on $^{nat}$Cr covers nearly the whole energy range, in agreement with previous data sets for $^{nat,50}$Cr \cite{herm,klein,coetz,baron} on distinct energy ranges shown in Figs.~\ref{48V_natCr}(a,b). 

Then, the sequential nucleon emission and particularly the $^{52}$Cr$(d,\alpha 2n)$ reaction provides the major contribution to $^{48}$V population by deuterons on $^{nat}$Cr, which was measured previously up to 50 MeV \cite{herm}. A suitable agreement with all experimental cross sections for deuteron-induced reactions on Cr isotopes by the present work analysis and TENDL-2017 is firstly met as well. A motivation of the similar calculated results could be the weak component of the DR pick-up $(d,\alpha)$ mechanism, pointed out in Sec.~\ref{DR} to be due to the scarce specific spectroscopic information as, e.g., experimental angular distributions, spectroscopic factors for residual levels populated by pick-up process, and transferred orbital momentum. 
Therefore, the description of the experimental $^{nat}$Cr$(d,x)^{48}$V and $^{50}$Cr$(d,x)^{48}$V excitation functions supports consistently the present model calculations with the assumption of a weak $(d,\alpha)$ pick-up component, while the above-discussed weight of the BF component has also been well decreased.

\subsubsection{The $(d,\alpha xn yp)$  reactions}

Population of $^{46,47}$Sc residual nuclei follows mainly the PE+CN statistical mechanisms (Fig.~\ref{4746Sc_natCr}), the BF enhancement brought by breakup nucleons being mostly weak. The only different case is that of $^{50}$Cr$(d,x)^{46}$Sc reaction in Fig.~\ref{4746Sc_natCr}(g), at incident energies above $\sim$50 MeV, with an odd-odd residual nucleus in the BU-neutron induced reaction $(n,p\alpha)$. 
However, the trend of the only available experimental data for reaction  $^{nat}$Cr$(d,x)^{46}$Sc in Fig.~\ref{4746Sc_natCr}(f) is given by the corresponding reaction on $^{52}$Cr. These data are well described by the present approach while the TENDL-2017 evaluation largely underestimates them. 

\begin{figure*} 
\resizebox{2.06\columnwidth}{!}{\includegraphics{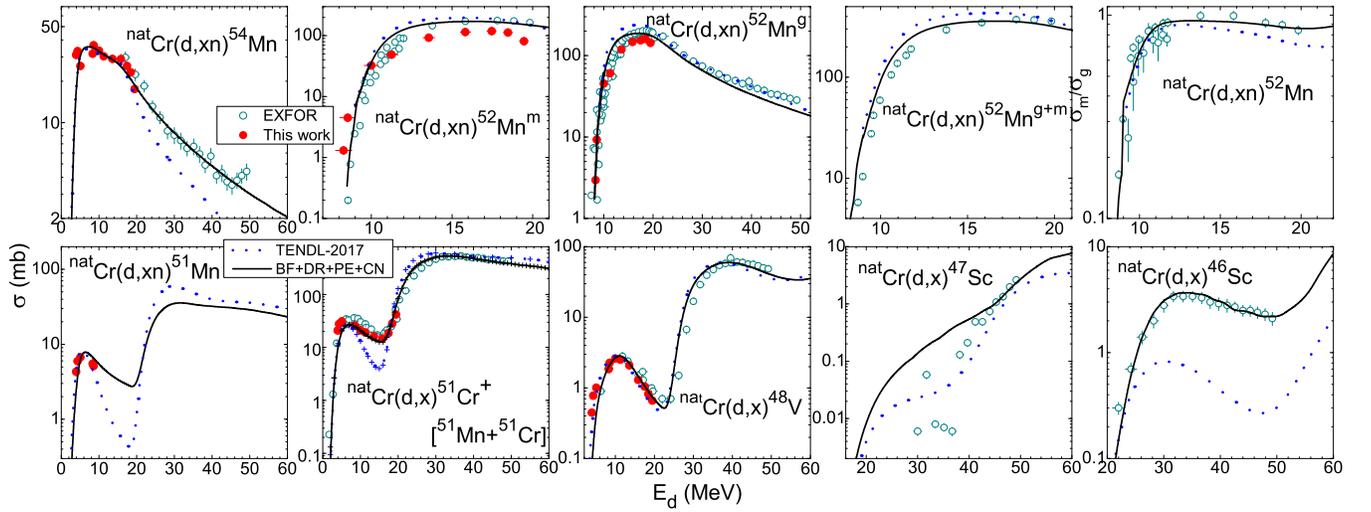}}
\caption{(Color online) Comparison of previous \cite{herm,ochiai,west,cheng,burg,cogn,klein,coetz,bisconti,baron} (open circles) and present measurements (solid circles), TENDL-2017 \cite{TENDL} predictions (dotted curves), and present calculations (solid curves) of cross sections for deuteron interactions with $^{nat}$Cr (see text).}
\label{natCr_fin}
\end{figure*}

On the other hand, the only available experimental data for reaction  $^{nat}$Cr$(d,x)^{47}$Sc [Fig.~\ref{4746Sc_natCr}(a)] are not reproduced at the incident energies lower than $\sim$38 MeV by either the present analysis or TENDL-2017.
Actually, it seems that some agreement is provided by the two model approaches at the energies where the $^{47}$Sc population is determined by the most-abundant isotope $^{52}$Cr, while the problems correspond to the incident energies where the main role is played by the $(d,2\alpha)$ and $(d,n2\alpha)$ reaction channels, for $^{53,54}$Cr target nuclei. However, the same channels are well described in the case of $^{46}$Sc population. 
Therefore, additional measurements of the former residual nucleus and deuteron energies of 30-40 MeV seems to be of interest as long as Sc isotopes are important for the development of therapeutic radio-pharmaceuticals \cite{Sc}. 

\section{CONCLUSIONS}
\label{Sum}

The activation cross sections for production of $^{51,52,54}$Mn, $^{51}$Cr, and $^{48}$V radioisotopes in deuteron-induced reactions on natural Cr were measured at deuteron energies up to 20 MeV, highly requested by the large-scale research projects \cite{iter,ifmif,nfs}. 
They are in good agreement with the previously reported experiments \cite{herm,ochiai,west,cheng,burg,cogn,klein,coetz,bisconti,baron} while all of them have been the object of an extended analysis from elastic scattering until the evaporation from fully equilibrated compound system. A particular attention has been given at the same time to breakup and direct reactions mechanisms. 

A detailed theoretical treatment of each reaction mechanism contribution has made possible a reliable understanding of the interaction process as well as accurate values of the calculated deuteron activation cross sections. Furthermore, the comparison of the experimental deuteron activation cross sections with both our model calculations and the corresponding TENDL-2017 evaluation supports the detailed theoretical treatment of deuteron interaction process, while the discrepancies between the measured data and corresponding TENDL-2017 evaluations have been explained as the result of overlooking the inelastic breakup enhancement, as well as of the inappropriate treatment of stripping and pick-up processes. 

This comparison particularly points out the importance of the new measured cross sections around the maximum of the $^{nat}$Cr$(d,xn)^{54}$Mn excitation function as well as the role of breakup and stripping mechanisms to provide the suitable description of these data.
The firstly measured cross sections for $^{nat}$Cr$(d,xn)^{51}$Mn reaction in the present work, in addition to only a couple of data sets already available for $^{50}$Cr$(d,n)^{51}$Mn \cite{cogn,klein}, play also a similar role. 
The case is similar for the newly measured data around the first maximum of the cumulative $^{nat}$Cr$(d,x)^{51}$Cr excitation function, which is related to the one for $^{50}$Cr.

However, while the associated theoretical models for stripping, pick-up, PE and CN are already settled, an increased attention should be paid to the theoretical description of the breakup mechanism, including its inelastic component. The recently increased interest on the theoretical analysis of the breakup components \cite{lei18,CDCC4,neoh,lei15} may lead eventually to the refinement of the deuteron breakup empirical parametrization and increased accuracy of the deuteron activation cross section calculations, well beyond reaction cross sections recommended most recently for high-priority elements still using data fit by various-order Pade approximations \cite{herm18}.

On the other hand, the improvement of the deuteron breakup description requires, beyond the increase of its own data basis, also complementary measurements of $(d,px)$ and $(n,x)$, as well as $(d,nx)$ and $(p,x)$ reaction cross sections for the same target nucleus, within corresponding incident-energy ranges.

\section{Acknowledgments}

This work has been partly supported by OP RDE, MEYS, Czech Republic under the projects SPIRAL2-CZ, CZ.02.1.01/0.0/0.0/16\_013/0001679, 
and Autoritatea Nationala pentru Cercetare Stiintifica (Project PN-16420102), within the framework of the EUROfusion Consortium and has received funding from the Euratom research and training programme 2014-2018 under grant agreement No 633053. The views and opinions expressed herein do not necessarily reflect those of the European Commission. 

\bibliography{CU10574proofs+}

\end{document}